\documentclass[aps,prl,amsfonts,amssymb,twocolumn,amsmath,preprintnumbers,nofootinbib,floatfix,superscriptaddress]{revtex4-1}%
\usepackage{siunitx}
\usepackage{graphicx}
\usepackage{mathrsfs}
\usepackage{bm}
\usepackage{amsmath}
\usepackage{dcolumn}
\usepackage{epstopdf}
\usepackage{dsfont}
\usepackage{amssymb}
\usepackage{tabularx}
\usepackage{array}
\usepackage{float}
\usepackage{color}
\usepackage{epstopdf}
\usepackage{mathrsfs}
\usepackage{multirow}
\usepackage[T1]{fontenc}
\usepackage[colorlinks,linkcolor=blue,anchorcolor=blue,citecolor=blue,urlcolor=blue]{hyperref}
\providecommand{\U}[1]{\protect\rule{.1in}{.1in}}

\newcommand{\vect}[1]{\boldsymbol{#1}}

\usepackage{helvet}
\newcommand\T{\rule{0pt}{2.5ex}}       
\newcommand\B{\rule[-2.5ex]{0pt}{0pt}} 

\usepackage{amsfonts}
\usepackage{mathrsfs}
\usepackage{amsmath}
\usepackage{color}
\usepackage{graphicx}
\usepackage{bm}
\usepackage{amssymb}
\usepackage{xspace}
\usepackage{epstopdf}
\usepackage{dcolumn}
\usepackage{longtable}
\usepackage{multirow}
\usepackage{float}
\usepackage{comment}

\usepackage{pifont}

\begin{document}

\title{Intrinsic gyrotropic magnetic current from Zeeman quantum geometry}
\author{Longjun Xiang}
\thanks{These authors contributed equally to this work.}
\affiliation{College of Physics and Optoelectronic Engineering, Shenzhen University, Shenzhen 518060, China}

\author{Jinxiong Jia}
\thanks{These authors contributed equally to this work.}
\affiliation{College of Physics and Optoelectronic Engineering, Shenzhen University, Shenzhen 518060, China}
\affiliation{Department of Physics, University of Science and Technology of China, Hefei, Anhui 230026, China}

\author{Fuming Xu}
\affiliation{College of Physics and Optoelectronic Engineering, Shenzhen University, Shenzhen 518060, China}
\affiliation{Quantum Science Center of Guangdong-Hongkong-Macao Greater Bay Area (Guangdong), Shenzhen 518045, China}

\author{Zhenhua Qiao}
\affiliation{Department of Physics, University of Science and Technology of China, Hefei, Anhui 230026, China}

\author{Jian Wang}
\email[]{jianwang@hku.hk}
\affiliation{College of Physics and Optoelectronic Engineering, Shenzhen University, Shenzhen 518060, China}
\affiliation{Quantum Science Center of Guangdong-Hongkong-Macao Greater Bay Area (Guangdong), Shenzhen 518045, China}
\affiliation{Department of Physics, The University of Hong Kong, Pokfulam Road, Hong Kong, China}

\begin{abstract}
Quantum geometric tensor (QGT),
which is usually obtained by evaluating the quantum distance between Bloch states
parametrized by momentum,
plays a key role in exploring the exotic responses of quantum materials.
Herein, we revisit the concept of QGT by further taking into account
the spin degree of freedom.
Besides the conventional QGT relating to momentum translation, 
we uncover a new QGT (termed Zeeman QGT) relating to momentum translation as well as spin rotation,
whose imaginary (real) part gives the Zeeman Berry curvature (quantum metric).
Notably, we show that these novel quantum geometric quantities
can drive an intrinsic gyrotropic magnetic current (IGMC)
in spin-orbit coupled materials
when the electron spin is steered by an oscillating magnetic field.
With symmetry analysis, we show that a wide range of materials can support the IGMC,
as illustrated by model calculations.
Finally, we discuss the experimental aspects of detecting the IGMC
driven by Zeeman QGT.
\end{abstract}

\maketitle

\noindent{\textit{\textcolor{blue}{Introduction.}}} ---
It is well known that the quantum wavefunction of electrons in quantum materials
can have a tremendous impact on its dynamics
\cite{XiaoD2010, MaQ2021, Amit2024, MaQ2023, Yan2020, JA2022, AG2020, ThermalNoise, Berry, BJYang2020}.
Very often, the information of wavefunction is encoded
into the quantum geometric tensor (QGT) \cite{Berry, BJYang2020, Provost, RCheng},
whose imaginary part and real part give the Berry curvature and the quantum metric, respectively.
The Berry curvature is indispensable to understanding
the responses of electrons in crystalline solids
under external fields, such as the integer quantum Hall effect \cite{TKNN},
the intrinsic anomalous Hall effect \cite{Nagaosa2010},
and the nonlinear Hall effects \cite{FuL2015, KTLaw, MaBCD, KKBCD, BCQ}.
Recently, the previously rarely explored quantum metric also received much attention,
particularly in achieving the intrinsic nonlinear Hall effects 
\cite{GaoYang2014, QMD1, QMD2, QMD21, QMD3, QMD4, QMD5, QMD6}.
Encouraged by this progress, an intriguing direction is to explore the novel
quantum geometry beyond Berry curvature and quantum metric \cite{SYXu2024}.
Fundamentally, the QGT is achieved by considering the infinitesimal variation of Bloch states parametrized by momentum.
It is natural to ask if the novel quantum geometry
can be developed by considering the infinitesimal variation
of Bloch states parametrized by other degrees of freedom (such as spin).

In this \textit{Letter}, we answer the question in the affirmative.
We generalize the concept of QGT by calculating the quantum distance
in the Hilbert space spanned by Bloch states parametrized by both crystal momentum and spin.
We show that along with the conventional QGT that is only related to the momentum translation,
a new QGT termed Zeeman QGT \cite{Zeeman} can be obtained,
which is related to momentum translation and spin rotation.
In analogy with the conventional QGT, the imaginary (real) part of the Zeeman QGT
is defined as Zeeman Berry curvature (quantum metric).
Importantly, with the response theory \cite{Sipe0, Sipe1},
the semiclassical theory \cite{GaoYang2014, semi1, semi2, Jia2024},
and the Green's function method \cite{Book, Daodao},
we show that these novel quantum geometric quantities can
lead to an intrinsic gyrotropic magnetic current (IGMC) \cite{Zhong2016},
where the Bloch electrons in spin-orbit coupled crystals
solely driven by an oscillating magnetic field
form a longitudinal or transverse current.
Further, by symmetry analysis, we show that the IGMC can be expected in a wide range of materials.
Guided by symmetry, we evaluate the IGMC with model Hamiltonians.
Finally, we discuss the experimental aspects of probing the Zeeman-QGT-driven IGMC.
Our \textit{Letter} not only establishes the Zeeman quantum geometry of Bloch electrons
but also reveals the IGMC as its \textit{geometric response function}.

\bigskip
\noindent{\textit{\textcolor{blue}{Zeeman QGT.}}} ---
Usually, the QGT in solid-state physics is obtained by evaluating the
quantum distance between \textit{two infinitesimal neighboring points}
of the Hilbert space spanned by the cell-periodic Bloch state $|u_{n\vect{k}}^\chi\rangle$,
which is the eigenstate of the periodic Hamiltonian
with spin-orbit coupling (SOC) and hence is parametrized by $\vect{k}$ for crystal momentum 
and $\chi$ for spin degrees of freedom.
In particular, we find \cite{Berry, BJYang2020, Provost, RCheng, sup}
\begin{align}
ds_1^2
\equiv
||U_{d\vect{k}}|u_{n\vect{k}}^\chi\rangle-|u_{n\vect{k}}^\chi\rangle||^2
:=
\sum_{m \neq n} g_{nm}^{\alpha\beta} dk_\alpha dk_\beta,
\label{quantumdis1}
\end{align}
where $U_{d\vect{k}} \equiv e^{-id\vect{k}\cdot\hat{\vect{r}}}$ represents an infinitesimal
momentum translation generated by the position operator
$\hat{r}_\alpha=i\partial/\partial k_\alpha \equiv i\partial_\alpha$ ($\hbar=1$),
and $g_{nm}^{\alpha\beta}$ is the QGT defined by \cite{JA2022}
\begin{align}
g^{\alpha\beta}_{nm} = r^\alpha_{nm} r^\beta_{mn}
&=
\mathcal{G}^{\alpha\beta}_{nm} - \frac{i}{2} \Omega_{nm}^{\alpha\beta}
,
\end{align}
where $r_{nm}^\alpha \equiv \langle u_{n\vect{k}}^\chi |i\partial_\alpha|u_{m\vect{k}}^\chi \rangle$ with $n \neq m$
is the interband Berry connection,
$\mathcal{G}_{nm}^{\alpha\beta}=(r^\alpha_{nm}r^\beta_{mn}+r^\beta_{nm}r^\alpha_{mn})/2$
is the symmetric (local) quantum metric and
$\Omega_{nm}^{\alpha\beta}=i(r^\alpha_{nm}r^\beta_{mn}-r^\beta_{nm}r^\alpha_{mn})$
is the anti-symmetric (local) Berry curvature.

We note that Eq.~(\ref{quantumdis1}) only considers the infinitesimal variation of momentum $\vect{k}$.
Besides that, $|u_{n\vect{k}}^\chi\rangle$ is also parametrized by $\chi$ for spin degrees of freedom,
which regulates that $|u_{n\vect{k}}^\chi\rangle$ is a superposed state of spin-up
$|\uparrow\rangle=[1,0]^T$ and spin-down $|\downarrow \rangle=[0,1]^T$
as explicitly given by 
$|u_{n\vect{k}}^\chi\rangle=\left[ |u_{n\vect{k}}^a\rangle, |u_{n\vect{k}}^b\rangle \right]^T
=
|u_{n\vect{k}}^a\rangle|\uparrow\rangle+|u_{n\vect{k}}^b\rangle|\downarrow\rangle$,
where $|u_{n\vect{k}}^{a/b}\rangle$ represents the spatial wavefunction.
As a result, after applying $U_{d\vect{k}}$ to $|u_{n\vect{k}}^\chi\rangle$,
we can further apply an infinitesimal spin rotation $U_{d\vect{\theta}}$
to $U_{d\vect{k}}|u_{n\vect{k}}^\chi\rangle$ and generalize the quantum distance as \cite{footnote1}
\begin{align}
ds_2^2
& \equiv
|| U_{d\vect{\theta}} U_{d\vect{k}}|u_{n\vect{k}}^\chi\rangle - |u_{n\vect{k}}^\chi\rangle ||^2,
\label{quantumdis2}
\end{align}
where $U_{d\vect{\theta}} \equiv e^{-id\vect{\theta}\cdot\hat{\vect{\sigma}}/2}$ 
is generated by the spin angular momentum operator $\hat{\vect{\sigma}}/2$ ($\hbar=1$).
When $d\vect{\theta}=0$, Eq.~(\ref{quantumdis2}) reduces to Eq.~(\ref{quantumdis1});
however, when $d\vect{\theta} \neq 0$, by expanding
$U_{d\vect{\theta}}U_{d\vect{k}}=(1-id\theta_\alpha\hat{\sigma}_\alpha/2+\cdots)(1+dk_\beta\partial_\beta+\cdots)
=1+dk_\beta\partial_\beta-id\theta_\alpha\hat{\sigma}_\alpha/2+\cdots$,
we arrive at \cite{sup}
\begin{align}
ds_2^2 
&=
\sum_{m \neq n}
g_{nm}^{\alpha\beta} dk_\alpha dk_\beta
+
\sum_{m}
\Sigma_{nm}^{\alpha\beta}
d\theta_\alpha d\theta_\beta/4
\nonumber \\
&+
\sum_{m \neq n}
\left(z_{mn}^{\beta\alpha} + z^{\beta\alpha}_{nm}\right)
d\theta_\alpha dk_\beta/2
.
\label{allterm}
\end{align}
As expected, in addition to the conventional QGT $g^{\alpha\beta}_{nm}$,
we further obtain $\Sigma^{\alpha\beta}_{nm}=\sigma^\alpha_{nm}\sigma^\beta_{mn}$
with
$\sigma^\alpha_{mn} \equiv \langle u_{m\vect{k}}^\chi |\hat{\sigma}_\alpha |u_{n\vect{k}}^\chi\rangle$
and $z_{nm}^{\alpha\beta}=r^\alpha_{nm}\sigma^\beta_{mn}$,
where the latter one is termed Zeeman QGT which we will focus on below \cite{secondterm}.

In analogy with the conventional QGT,
by defining $z_{nm}^{\alpha\beta}=\mathcal{Q}_{nm}^{\alpha\beta}-\frac{i}{2}\mathcal{Z}^{\alpha\beta}_{nm}$,
we obtain the (local) Zeeman quantum metric ($\mathcal{Q}^{\alpha\beta}_{nm}$)
and Zeeman Berry curvature ($\mathcal{Z}^{\alpha\beta}_{nm}$)
\begin{align}
\mathcal{Q}_{nm}^{\alpha\beta} &= (r^\alpha_{nm} \sigma^\beta_{mn} + r^\alpha_{mn} \sigma^\beta_{nm})/2,
\label{ZQM}
\\
\mathcal{Z}_{nm}^{\alpha\beta} &= i\left( r^\alpha_{nm} \sigma^\beta_{mn} - r^\alpha_{mn} \sigma^\beta_{nm} \right).
\label{ZBC}
\end{align}
Despite the similar terminology, the Zeeman quantum metric (Berry curvature)
is no longer symmetric (anti-symmetric) about $\alpha$ and $\beta$. 
However, by further defining 
$\mathcal{Q}^{A/S;\alpha\beta}_{nm} \equiv (\mathcal{Q}^{\alpha\beta}_{nm} \mp \mathcal{Q}^{\beta\alpha}_{nm})/2$
and
$\mathcal{Z}^{A/S;\alpha\beta}_{nm} \equiv (\mathcal{Z}^{\alpha\beta}_{nm} \mp \mathcal{Z}^{\beta\alpha}_{nm})/2$,
the symmetric Zeeman quantum metric ($\mathcal{Q}^{S;\alpha\beta}_{nm}$) and
the anti-symmetric Zeeman Berry curvature ($\mathcal{Z}^{A;\alpha\beta}_{nm}$)
similar to that in the conventional QGT can be obtained,
while the anti-symmetric quantum metric ($\mathcal{Q}^{A;\alpha\beta}_{nm}$)
and the symmetric Zeeman Berry curvature ($\mathcal{Z}^{S;\alpha\beta}_{nm}$) 
have no counterpart in the conventional QGT.
In vector notation, the anti-symmetric Zeeman Berry curvature
can be further expressed as $\vect{\mathcal{Z}}^A_n = \nabla_{\vect{k}} \times \vect{\sigma}_n/2$
like the conventional Berry curvature
$\vect{\Omega}_n = \nabla_{\vect{k}} \times \vect{\mathcal{A}}_n$ \cite{XiaoD2010}.

Furthermore, we remark that the Zeeman quantum metric (Berry curvature)
shows a completely different symmetry transformation
compared to its counterpart in the conventional QGT.
For instance, under time reversal ($\mathcal{T}$) operation, we find
that $\Omega_{nm}^{\alpha\beta}$ is $\mathcal{T}$-odd while $\mathcal{Z}_{nm}^{\alpha\beta}$ is $\mathcal{T}$-even
since $\mathcal{T}r^\alpha_{nm}=r^\alpha_{mn}$ but $\mathcal{T}\sigma^\alpha_{nm}=-\sigma^\alpha_{mn}$;
under inversion ($\mathcal{P}$) operation,
both $\mathcal{G}_{nm}^{\alpha\beta}$ and $\Omega_{nm}^{\alpha\beta}$
are $\mathcal{P}$-even while both $\mathcal{Q}_{nm}^{\alpha\beta}$ and $\mathcal{Z}_{nm}^{\alpha\beta}$
are $\mathcal{P}$-odd since $\mathcal{P}r^\alpha_{nm}=-r^\alpha_{nm}$ but $\mathcal{P}\sigma^\alpha_{nm}=\sigma^\alpha_{nm}$.
In Table \ref{tab0}, a full comparison between these two QGTs
for $\mathcal{P}$, $\mathcal{T}$, and $\mathcal{P}\mathcal{T}$ is listed.

Finally, we remark that the Zeeman QGT is trivial in the systems without SOC,
such as in a ferromagnet.
Specifically, for a ferromagnet polarized along $z$ direction,
where $|u_{n\vect{k}}^\chi\rangle=[|u_{n\vect{k}}^a\rangle,0]^T$,
we find
$\sigma^\beta_{mn}=\langle u_{m\vect{k}}^\chi|\hat{\sigma}_\beta|u_{n\vect{k}}^\chi\rangle
=
\langle u_{m\vect{k}}^a|u_{n\vect{k}}^a\rangle [1, 0] \hat{\sigma}_\beta [1, 0]^T=0
$
for $m \neq n$ since
$\langle u_{mk}^\chi|u_{n\vect{k}}^\chi\rangle=\langle u^a_{m\vect{k}}|u^a_{n\vect{k}}\rangle=\delta_{mn}$.
However, the Zeeman QGT can be nontrivial when the SOC is taken into account,
where $\sigma^\beta_{mn} \neq 0$ in general since 
$\langle u_{m\vect{k}}^\chi|u_{n\vect{k}}^\chi\rangle
=
\langle u_{m\vect{k}}^a|u_{n\vect{k}}^a\rangle
+
\langle u_{m\vect{k}}^b|u_{n\vect{k}}^b\rangle
=
\delta_{mn}
$.

\begin{center}
\begin{table}[t!]
\caption{\label{tab0}
The transformation of quantum metric $(\mathcal{G}_{nm}^{\alpha\beta})$,
Berry curvature $(\Omega_{nm}^{\alpha\beta})$, Zeeman quantum metric ($\mathcal{Q}_{nm}^{\alpha\beta}$)
and Zeeman Berry curvature $(\mathcal{Z}_{nm}^{\alpha\beta})$ under inversion $(\mathcal{P})$,
time reversal $(\mathcal{T})$, and $\mathcal{P}\mathcal{T}$ operations.
Here $+ (-)$ indicates the even (odd) property under the corresponding operation.}
\begin{tabular}{ p{2.5cm} c c c c}
\hline
\hline
\\
                        \ \ & \ \ \ $\mathcal{G}^{\alpha\beta}_{nm}$ \ \ \  & \ \ \ $\Omega_{nm}^{\alpha\beta}$ \ \ \
                            & \ \ \ $\mathcal{Q}^{\alpha\beta}_{nm}$ \ \ \  & \ \ \ $\mathcal{Z}^{\alpha\beta}_{nm}$ \ \ \ \T\B
\\
[0.05em]
\hline
$\mathcal{P}$ \ \ & $+$ & $+$ & $-$ & $-$
\\
[0.05em]
$\mathcal{T}$ \ \ & $+$ & $-$ & $-$ & $+$
\\
[0.05em]
$\mathcal{P}\mathcal{T}$ \ \ & $+$ & $-$ & $+$ & $-$
\\
[0.05em]
\hline
\hline
\end{tabular}
\end{table}
\end{center}

\bigskip
\noindent{\textit{\textcolor{blue}{IGMC from Zeeman QGT.}}} ---
It has been known that both $\mathcal{G}_{nm}^{\alpha\beta}$ and $\Omega_{nm}^{\alpha\beta}$ can be probed by
detecting the current response under external fields, here we show that
this is also the case for both $\mathcal{Q}_{nm}^{\alpha\beta}$ and $\mathcal{Z}_{nm}^{\alpha\beta}$.
Since spin is coupled to the magnetic field \cite{orbitalB},
we consider the current response driven by a uniform oscillating magnetic field $\vect{B}(t)$
with a perturbed Hamiltonian $H'=-g\mu_B\vect{B}(t)\cdot\vect{\sigma}$,
where $g$ is the $g$-factor, $\mu_B$ the Bohr magneton,
and $\vect{B}(t)=\frac{1}{2} \vect{B} \sum_{\omega_1} e^{-i \omega_1 t}$ with $\omega_1=\pm\omega$.
Following the standard perturbation approach \cite{Sipe0, Sipe1},
by iteratively solving the quantum Liouville equation,
the density matrix element to the first order of $\vect{B}$ field is given by
\cite{sup} (we set $e=\hbar=1$)
\begin{align}
\rho_{mn}^{(1)}=
-\frac{g \mu_B }{2} \sum_{\omega_1}
\frac{f_{nm}{\sigma}^\alpha_{mn} B^\alpha e^{-i \omega_1 t}}{\omega_1-\epsilon_{mn}+i\eta},
\label{rho1}
\end{align}
where $f_{nm} \equiv f_n-f_m$ with $f_n$ the equilibrium Fermi distribution function,
$\eta$ is an infinitesimal quantity,
and $\sigma^\alpha_{mn}$ is the matrix element of $\hat{\sigma}_\alpha$ appeared in 
$\mathcal{Q}^{\alpha\beta}_{nm}$ and $\mathcal{Z}_{nm}^{\alpha\beta}$.
Note that the extrinsic part of the density matrix has been neglected
and the summation over the repeated Greek alphabets is assumed here and hereafter.

With Eq.~(\ref{rho1}), the intrinsic current density, 
defined by $J_\alpha = \int_k v^\alpha_{nm}\rho^{(1)}_{mn}$,
is found to be
\begin{align}
J_\alpha
=
-\frac{g \mu_B}{2}
\sum_{mn}
\sum_{\omega_1}
\int_k
\frac{f_{nm} v^\alpha_{nm} \sigma^\beta_{mn} B^\beta e^{-i \omega_1 t}}{\omega_1-\epsilon_{mn}+i\eta},
\label{Jalpha1}
\end{align}
where $\int_k = \int d\vect{k}/(2\pi)^d$ with $d$ the spatial dimension and
$v^\alpha_{nm}$ is the interband velocity matrix element.
Further, by assuming $\hbar\omega \ll \epsilon_{mn}$,
we can focus on the non-resonant response. As a result,
by using $v^\alpha_{nm}=i\epsilon_{nm}r^\alpha_{nm} (m \neq n)$,
Eq.~(\ref{Jalpha1}) can be recast into
\begin{align}
J_\alpha
&=
-g\mu_B\sum_{nm}\int_k f_n \mathcal{Z}^{\alpha\beta}_{nm} B^\beta \cos\omega t
+
\partial_t P_\alpha (t),
\label{Jalpha2}
\\
P_{\alpha}
&=
-g\mu_B\sum_{n} \int_k f_n [\mathcal{N}^{\alpha\beta}_{n}\cos\omega t+ \mathcal{F}^{\alpha\beta}_n\sin\omega t] B^\beta,
\label{Pt}
\end{align}
where $P_\alpha(t)$ with 
$\mathcal{N}^{\alpha\beta}_n =
-\sum_m 
2\mathcal{Q}_{nm}^{\alpha\beta}/\epsilon_{mn}
$
and
$\mathcal{F}^{\alpha\beta}_n
=
-\sum_m
\omega \mathcal{Z}_{nm}^{\alpha\beta}/\epsilon_{mn}^2
$
is the AC charge polarization induced by the magnetic field.

As expected, we find that both the Zeeman Berry curvature $(\mathcal{Z}_{nm}^{\alpha\beta})$
and the Zeeman quantum metric $(\mathcal{Q}_{nm}^{\alpha\beta})$
show up in the current response.
Specifically, the first term of Eq.~(\ref{Jalpha2}) arising from $\mathcal{Z}_{nm}^{\alpha\beta}$
gives the conduction current while the second term of Eq.~(\ref{Jalpha2}) arising from both
$\mathcal{Q}_{nm}^{\alpha\beta}$ and $\mathcal{Z}_{nm}^{\alpha\beta}$
corresponds to the displacement current \cite{displacement}.
The conduction current induced by an oscillating magnetic field
was studied in Ref.~[\onlinecite{Zhong2016}],
which is termed gyrotropic magnetic current in analogy with
the gyrotropic current achieved by rectifying the optical fields \cite{Landau, Ginzburg}.
However, we wish to remark that the current response discussed in Ref.~[\onlinecite{Zhong2016}]
depends on the relaxation time $\tau$ and hence represents an extrinsic effect.
Conversely, the conduction (displacement) current response studied here is free of $\tau$ and hence 
is termed conduction (displacement) intrinsic gyrotropic magnetic current (IGMC).
Note that the intrinsic current response is fully determined by the information from Bloch states
and can occur in moderately dirty materials \cite{Nagaosa2006}.

Next, by defining $J_\alpha \equiv -g \mu_B \sigma^{C/D}_{\alpha\beta} B^\beta (t)$
with $\sigma^C_{\alpha\beta}$ ($\sigma^D_{\alpha\beta}$) denoting the conduction (displacement)
IGMC conductivity \cite{footnote2}, 
then from Eqs. (\ref{Jalpha2}-\ref{Pt}) we obtain 
\begin{align}
\sigma_{\alpha\beta}^C
&\equiv
\sum_{nm} \int_k f_n \mathcal{Z}^{\alpha\beta}_{nm},
\label{Ccurrent}
\\
\sigma_{\alpha\beta}^D
&\equiv
\sum_{nm} \int_k f_n \frac{2\omega}{\epsilon_{mn}} \mathcal{Q}^{\alpha\beta}_{nm},
\label{Dcurrent}
\end{align}
where we have ignored the second term in Eq.~(\ref{Pt}) since it is proprotional to $\omega^2$ \cite{displacement}.
Several remarks for these two conductivities are given in order.
First, we remark that Eqs.~(\ref{Ccurrent}-\ref{Dcurrent})
can also be obtained with the semiclassical theory and the Green's function method,
as reproduced in the Supplemental Material~\cite{sup}.
Second, we note that Eqs.~(\ref{Ccurrent}-\ref{Dcurrent})
can include a transverse contribution,
which leads to the Hall IGMC.
Third, despite their formal similarity, we wish to mention that
Eq.~(\ref{Dcurrent}) shows a Fermi-sea property
like the intrinsic spin Hall conductivity \cite{ZhangScience, Sinova2004}
while Eq.~(\ref{Ccurrent}) in fact stands for a Fermi-surface property,
the same as the conduction current driven by the electric field.
Explicitly, by using $\sum_{m} \mathcal{Z}_{nm}^{\alpha\beta}=\partial_\alpha \sigma^\beta_{nn}$,
Eq.~(\ref{Ccurrent}) can be rewritten as
\begin{align}
\sigma_{\alpha\beta}^C
=
\sum_{n} \int_k f_n \partial_\alpha \sigma^\beta_{nn}
=
-\sum_{n} \int_k \frac{\partial f_n}{\partial \epsilon_n} v_n^\alpha \sigma^\beta_{nn},
\label{Fermisurf}
\end{align}
which indicates that $\sigma^C_{\alpha\beta}$ vanishes when the chemical potential
is located in the gap at zero temperature.
Finally, Eqs.~(\ref{Ccurrent}-\ref{Dcurrent}), together with Eqs.~(\ref{ZQM}-\ref{ZBC}),
are the main results of our \textit{Letter} and can be used to evaluate the IGMC in realistic materials
when combined with the first-principles calculations \cite{abinitio} and symmetry analysis.

\begin{figure}[t!]
\includegraphics[width=0.95\columnwidth]{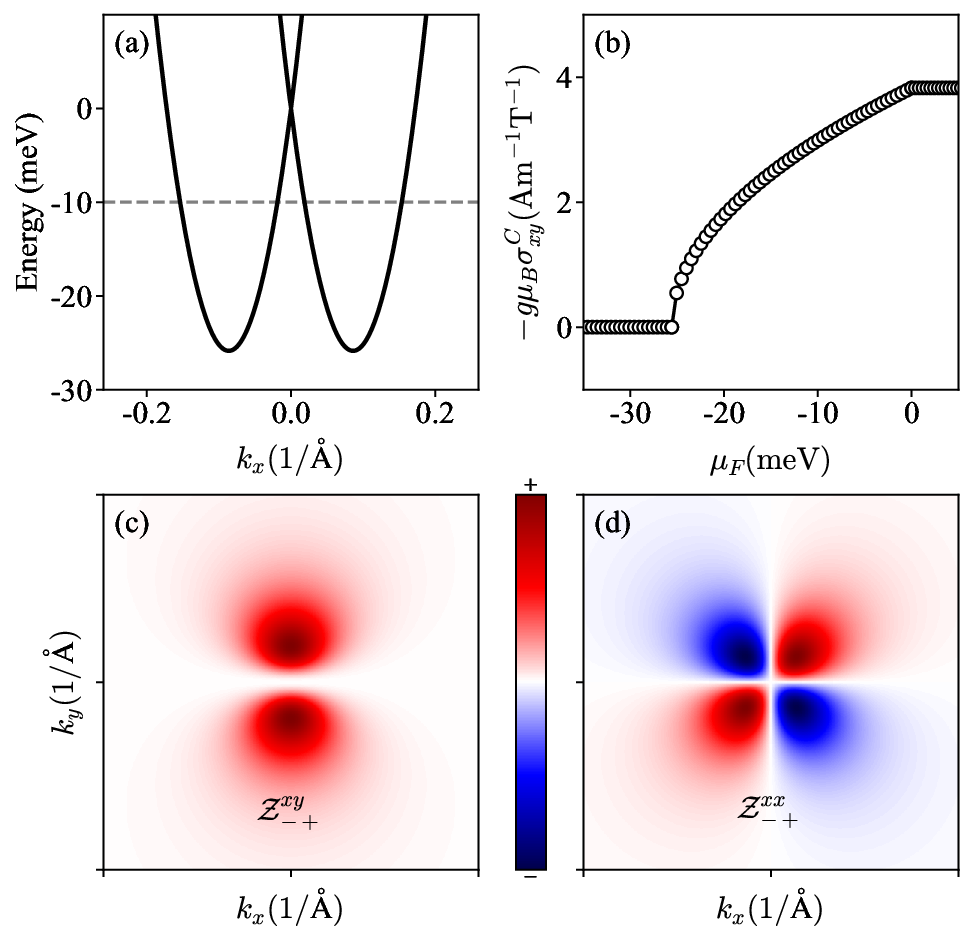}
\caption{
(a) Band dispersions for Eq.~(\ref{Rashba}).
The horizontal dashed line indicates the Fermi level $\mu_F$.
(b) The dependence of $\sigma_{xy}^C$ on $\mu_F$ with $\lambda_R=0.6 \mathrm{eV}\cdot\mathrm{\mathring{A}}$
and $m=1.1m_e$, where $m_e$ is the mass of the free electron.
The $\vect{k}$-resolved distribution for the Zeeman Berry curvature
(c) $\mathcal{Z}_{-+}^{xy}$ and (d) $\mathcal{Z}_{-+}^{xx}$.
}
\label{FIG1}
\end{figure}

\bigskip
\noindent{\textit{\textcolor{blue}{Symmetry analysis.}}} ---
Since $\sigma_{\alpha\beta}^C$ $(\sigma_{\alpha\beta}^D)$ is decided by
$\mathcal{Z}^{\alpha\beta}_{nm}$ $(\mathcal{Q}_{nm}^{\alpha\beta})$,
the same symmetry constraint under $\mathcal{P}$, $\mathcal{T}$,
and $\mathcal{P}\mathcal{T}$ can be derived \cite{evenodd}.
However, to obtain the complete constraints of $\sigma^{C/D}_{\alpha\beta}$
from the magnetic point groups (MPGs),
we can resort to the Neumann's principle \cite{Neumann}:
\begin{align}
\sigma^{C/D}_{\alpha \beta} &= \eta_T \text{det}(R) R_{\alpha \alpha '} R_{\beta \beta '} \sigma^{C/D}_{\alpha ' \beta '},
\label{sigmaCD}
\end{align}
where $R$ represents a point group operation,
$\text{det}(R)$ is responsible for the pseudo nature of $\sigma^{C/D}_{\alpha \beta}$,
and $\eta_T=+1(-1)$ for $R$ $(R\mathcal{T})$.
Eq.~(\ref{sigmaCD}) has been implemented into the Bilbao Crystallographic Server \cite{Bilbao}.
As a result, by defining a suitable Jahn's notation \cite{Jahn} $\mathrm{eV^2}$ ($\mathrm{aeV^2}$)
for $\sigma_{\alpha\beta}^{C}$ ($\sigma_{\alpha\beta}^{D}$),
all the allowed MPGs for $\sigma_{\alpha\beta}^{C}$ ($\sigma_{\alpha\beta}^{D}$)
can be found, as classified in TABLE III of the Supplementary Material \cite{sup},
where the specific tensor shape of $\sigma^{C}_{\alpha\beta}$ ($\sigma^{D}_{\alpha\beta}$) for these MPGs
is enumerated in TABLE I (II).
We find that both $\sigma^C_{\alpha\beta}$ and $\sigma^D_{\alpha\beta}$
are supported by many MPGs,
which correspond to a large number of realistic materials.
In addition, we note that some MPGs can only support the Hall IGMC, see the third column of
TABLE III of the Supplementary Material \cite{sup}.
Further, for $\sigma^C_{\alpha\beta}$, we find that
in $4mm$, $4mm1'$, $4'm'm$, $4m'm'$ $3m$, $3m1'$, $6mm$, $6mm1'$, $6'mm'$, and $6m'm'$,
there is only one independent element $\sigma_{xy}^C=-\sigma_{yx}^C$
as only contributed by $\vect{\mathcal{Z}}_n^A$.
Guided by symmetry, next we evaluate the IGMC with spin-orbit coupled model Hamiltonians.

\bigskip
\noindent{\textit{\textcolor{blue}{Rashba model.}}} ---
We first consider the 2D Rashba model with MPG $4mm1'$ \cite{Rashba}
\begin{align}
H=k^2/2m + \lambda_R(k_y\sigma_x-k_x\sigma_y),
\label{Rashba}
\end{align}
where $\sigma_\alpha$ is the Pauli matrix for spin and $k^2=k_x^2+k_y^2$.
The band dispersions of this model are $\epsilon_{\pm}=k^2/2m \pm \lambda_R k$
with $+ (-)$ the conduction (valence) band, as shown in Fig.~\ref{FIG1}(a).
This model preserves $\mathcal{T}$ symmetry ($1'$) and can only support the conduction IGMC,
see TABLE III of the Supplementary Material \cite{sup}.
As discussed above, for MPG $4mm1'$, we have only one independent element $\sigma^C_{xy}=-\sigma^C_{yx}$,
as can be checked by directly evaluating the Zeeman Berry curvature of Eq.~(\ref{Rashba}).
Explicitly, we find $\mathcal{Z}^{xy}_{\pm\mp} = \mp k_y^2/k^3$, $\mathcal{Z}^{yx}_{\pm\mp} = \pm k_x^2/k^3$,
$\mathcal{Z}^{xx}_{\pm\mp} = \mp k_x k_y/k^3=-\mathcal{Z}^{yy}_{\pm\mp}$,
where $\mathcal{Z}^{xy}_{-+}$ and $\mathcal{Z}^{xx}_{-+}$ are displayed in Fig.~\ref{FIG1}(c-d).
Then by Eq.~(\ref{Ccurrent}), $\sigma^C_{xy}$
at zero temperature is evaluated as
\begin{align}
\sigma_{xy}^{C}
&=
\begin{cases}
\frac{e\sqrt{\lambda_R^2m^2+2\hbar^2m\mu_F}}{2\pi\hbar^3}, \mu_F \in [-\frac{\lambda_R^2m}{2\hbar^2},0) \\
\frac{e\lambda_Rm}{2\pi\hbar^3}, \mu_F \in [0, +\infty)
\end{cases},
\end{align}
where $e$ and $\hbar$ are restored by dimension analysis
and $\mu_F$ is the chemical potential.
In Fig.~\ref{FIG1}(b), the dependence of $\sigma^C_{xy}$ on $\mu_F$ 
is plotted. Below the band crossing point,
we find that $\sigma^C_{xy}$ decreases to zero when $\mu_F$ approaches the band edge;
above the band crossing point,
$\sigma^C_{xy}$ displays a constant behavior on $\mu_F$,
where the magnitude of $\sigma^C_{xy}$
is decided by the Rashba constant $\lambda_R$ and the effective electron mass $m$.

\begin{figure}[t!]
\includegraphics[width=0.95\columnwidth]{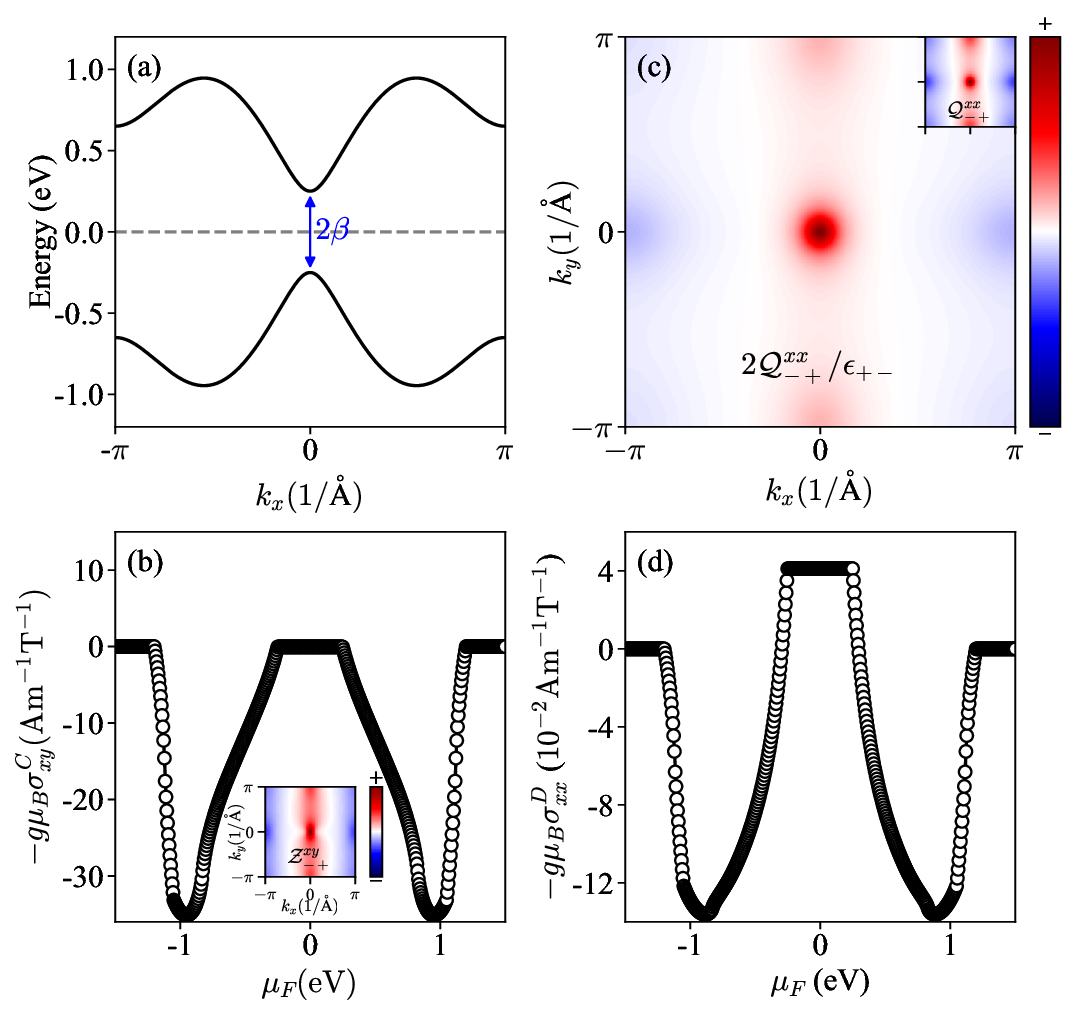}
\caption{
(a) Band dispersions for Eq.~(\ref{Dirac1}).
(b) The conduction Hall IGMC conductivity for Eq.~(\ref{Dirac1}).
Here the inset shows the contributed Zeeman Berry curvature.
(c) The band-resolved and bare (inset) Zeeman quantum metric for Eq.~(\ref{Dirac1}).
(d) The displacement longitudinal IGMC conductivity.
Parameters: $\beta=0.25 \mathrm{eV}$, $a=1 \mathrm{\mathring{A}}$, $v_F=0.825 \mathrm{eV}\cdot\mathrm{\mathring{A}}$,
$M=0.1 \mathrm{eV}\cdot\mathrm{\mathring{A}}^2$, and $\omega=10^{12} \mathrm{Hz}$.
}
\label{FIG2}
\end{figure}

\bigskip
\noindent{\textit{\textcolor{blue}{Dirac model.}}} ---
We next consider the Dirac model on a square lattice with MPG $4m'm'$,
whose Bloch Hamiltonian is given by
\begin{align}
\bar{H}
&=
v_F \frac{\sin(ak_y)}{a}\sigma_x - v_F\frac{\sin(ak_x)}{a}\sigma_y + \beta \sigma_z
\nonumber \\
&+
\frac{2M}{a^2}\left[2-\cos(ak_x)-\cos(ak_y)\right]\sigma_z 
\label{Dirac1},
\end{align}
where $\sigma_\alpha$ is also the Pauli matrix for spin,
$a$ is the lattice constant, $v_F$ is the Fermi velocity and $\beta$ controls the gap.
This model is obtained by projecting the continuum Dirac model
$H=v_F(k_y\sigma_x-k_x\sigma_y)+\beta\sigma_z$ \cite{Diracmodel} into a square lattice,
where the last term in Eq.~(\ref{Dirac1}) with $M \neq 0$ stands for a quadratic correction.
Eq.~(\ref{Dirac1}) breaks $\mathcal{P}$, $\mathcal{T}$, and $\mathcal{P}\mathcal{T}$ symmetries,
therefore, both the conduction and displacement IGMCs can be allowed in terms of TABLE III
of the Supplementary Material \cite{sup}.
By checking the tensor shape of $\sigma_{\alpha\beta}^{C}$ ($\sigma^D_{\alpha\beta}$)
for MPG $4m'm'$ listed in the Supplementary Material \cite{sup},
we find $\sigma^C_{xy}=-\sigma^C_{yx}$ ($\sigma^D_{xx}=\sigma^D_{yy}$).

In Fig.~\ref{FIG2}(a), the band dispersion for Eq.~(\ref{Dirac1}) is plotted,
which shows a band gap $2\beta$ near $\Gamma$ point.
In Fig.~\ref{FIG2}(b), $\sigma^C_{xy}$
for Eq.~(\ref{Dirac1}) is displayed, where the integrand
$\mathcal{Z}^{xy}_{\pm\mp}$ is also shown, see the inset of Fig.~\ref{FIG2}(b).
Further, $\sigma^{D}_{xx}$ for Eq.~(\ref{Dirac1})
is presented in Fig.~\ref{FIG2}(d).
In stark contrast with $\sigma^C_{xy}$ that vanishes in the gap,
an in-gap $\sigma^D_{xx}$ is observed \cite{displacement}.
Interestingly, we note that this in-gap conductivity arises from the band-resolved Zeeman quantum metric
$2\mathcal{Q}_{-+}^{xx}/\epsilon_{+-}$, which is localized around $\Gamma$ point,
as shown in Fig.~\ref{FIG2}(c),
instead of the bare Zeeman quantum metric $\mathcal{Q}^{xx}_{-+}$ distributed in the whole Brillouin zone,
see the inset of Fig.~\ref{FIG2}(c).
In addition, both $\sigma^{C}_{xy}$ and $\sigma^D_{xx}$ peak near $\mu_F=\pm 1 \mathrm{eV}$.

\bigskip
\noindent{\textit{\textcolor{blue}{Summary and discussion.}}} ---
In summary, we establish the Zeeman QGT by revisiting the quantum distance between
Bloch states particularly with the consideration of spin rotation.
Importantly, we show that the Zeeman QGT can drive an IGMC. 
To facilitate the search for materials exhibiting IGMC and thereby enable the detection of QGT,
a complete symmetry classification for the IGMC is tabulated.
From Fig.~\ref{FIG2}(b), by taking the lateral size $\sim 100 \mathrm{\mu m}$
and the resistance $\sim 10^3 \mathrm{\Omega}$ for the Hall bar \cite{KTLaw},
and by applying a small field $B=0.001 \mathrm{T}$ with a low driving frequency 
(such as $\omega=1 \sim 10^2 \mathrm{Hz}$),
we find that the conduction IGMC Hall voltage is $\sim 1 \mathrm{mV}$,
which can be accessible experimentally,
such as in 2D van der Waals heterostructures \cite{vdw} and
2D transition metal dichalcogenides \cite{TMDC}.
However, to detect the displacement IGMC,
a high-frequency (such as THz) magnetic field is necessary to obtain
an observable signal ($0.01 \mathrm{mV}$).

Although the IGMC is solely driven by the oscillating magnetic field,
some competing effects may appear in experiments.
First, the magnetic field through the minimal coupling can
induce a longitudinal chiral magnetic current \cite{CME1, CME2, CME3},
but it can only be expected in 3D Weyl semimetals \cite{Burkov2012, Son2012, Tewari2013}.
Indeed, by considering a two-dimensional system probed by an in-plane magnetic field,
the orbital effects can be safely neglected \cite{orbitalB}.
Second, the extrinsic gyrotropic magnetic current,
which contains the mechanism studied in Ref.~[\onlinecite{Zhong2016}]
and the possible side-jump and skew scattering mechanisms \cite{Nagaosa2010},
generally can be distinguished by investigating the scaling relation
and the symmetry restrictions of the conductivity tensor \cite{QMD21, scaling0, scaling, scaling1, scaling2},
which needs further theoretical efforts.
Third, the possible optical currents from the induced electric field,
which are also related to quantum geometry and are very interesting \cite{resonant1, resonant2, resonant3},
can be separated from the displacement IGMC by carefully choosing the material and external parameters,
as discussed in Supplementary Material \cite{sup}.

We close by noting that besides the conduction and displacement IGMCs,
the Zeeman QGT can also play a role in other spin-dependent dynamics,
such as spin polarization under electric field \cite{XiaoCPRL2022, XiaoPRL, XiaoarXiv},
which displays the potential capability of Zeeman QGT
and in turn, offers more response functions to probe the novel Zeeman quantum geometry.
Beyond the transport responses, the proposed Zeeman quantum geometry
may also appear in the spectroscopic responses (such as circular dichroism)
\cite{solid1, solid2, atom1, atom2, atom3, atom4} under light irradiation,
which will be explored in the future.

\bigskip
\noindent{{\bf Acknowledgments}} ---
This work was supported by the National Natural Science Foundation of China
(Grant Nos. 12034014 and 12174262).



\begin{thebibliography}{00}
\bibitem{Zeeman}
We use "Zeeman" since the spin degree of freedom
can be steered by the magnetic field through Zeeman coupling.
\bibitem{footnote1}
Similar to the two infinitesimal neighboring points
$U_{d\vect{k}}|u_{n\vect{k}}^\chi\rangle$ and $|u_{n\vect{k}}^\chi\rangle$
in Eq.~(\ref{quantumdis1}), the two points
$U_{d\vect{\theta}}U_{d\vect{k}}|u_{n\vect{k}}^\chi\rangle$ and $|u_{n\vect{k}}^\chi\rangle$
in Eq.~(\ref{quantumdis2}) are still infinitely close to each other
since both $U_{d\vect{k}}$ and $U_{d\vect{\theta}}$ are infinitesimal variations.
Additionally, we remark that $U_{d\vect{k}}|u_{n\vect{k}}^\chi\rangle=|u_{n\vect{k}+d\vect{k}}^\chi\rangle$
since $\vect{k}$ is a good quantum number but $U_{d\vect{\theta}}$ has not such a property
since $\chi$ for spin degrees of freedom is not a good quantum number due to SOC.
\bibitem{secondterm}
Note that $\Sigma^{\alpha\beta}_{nm}$ can also play a key role in spin-dependent dynamics,
such as in spin polarization induced by the magnetic field,
which will be explored in the future.
\bibitem{orbitalB}
Besides the spin degree of freedom $\vect{\sigma}$,
we note that the orbital degree of freedom 
can be probed by the magnetic field
through the minimal coupling $\vect{B}\cdot(\vect{r}\times\vect{v})$ \cite{Zhong2016},
but it is described by the conventional QGT \cite{GaoYang2014}.
However, by considering a two-dimension system pumped by an in-plane magnetic field,
the orbital effects can be neglected.
\bibitem{evenodd}
Namely, $\sigma_{\alpha\beta}^C$ $(\sigma_{\alpha\beta}^D)$
is a tensor of
$\mathcal{P}$-odd ($\mathcal{P}$-odd),
$\mathcal{T}$-even ($\mathcal{T}$-odd),
and $\mathcal{P}\mathcal{T}$-odd ($\mathcal{P}\mathcal{T}$-even).
\end{thebibliography}
\end{document}